\documentclass[onecolumn,draftclsnofoot]{IEEEtran}\usepackage[]{graphicx}\usepackage[]{color}
\makeatletter
\def\maxwidth{ %
  \ifdim\Gin@nat@width>\linewidth
    \linewidth
  \else
    \Gin@nat@width
  \fi
}
\makeatother

\definecolor{fgcolor}{rgb}{0.345, 0.345, 0.345}
\newcommand{\hlnum}[1]{\textcolor[rgb]{0.686,0.059,0.569}{#1}}%
\newcommand{\hlstr}[1]{\textcolor[rgb]{0.192,0.494,0.8}{#1}}%
\newcommand{\hlcom}[1]{\textcolor[rgb]{0.678,0.584,0.686}{\textit{#1}}}%
\newcommand{\hlopt}[1]{\textcolor[rgb]{0,0,0}{#1}}%
\newcommand{\hlstd}[1]{\textcolor[rgb]{0.345,0.345,0.345}{#1}}%
\newcommand{\hlkwa}[1]{\textcolor[rgb]{0.161,0.373,0.58}{\textbf{#1}}}%
\newcommand{\hlkwb}[1]{\textcolor[rgb]{0.69,0.353,0.396}{#1}}%
\newcommand{\hlkwc}[1]{\textcolor[rgb]{0.333,0.667,0.333}{#1}}%
\newcommand{\hlkwd}[1]{\textcolor[rgb]{0.737,0.353,0.396}{\textbf{#1}}}%

\usepackage{framed}
\makeatletter
\newenvironment{kframe}{%
 \def\at@end@of@kframe{}%
 \ifinner\ifhmode%
  \def\at@end@of@kframe{\end{minipage}}%
  \begin{minipage}{\columnwidth}%
 \fi\fi%
 \def\FrameCommand##1{\hskip\@totalleftmargin \hskip-\fboxsep
 \colorbox{shadecolor}{##1}\hskip-\fboxsep
     \hskip-\linewidth \hskip-\@totalleftmargin \hskip\columnwidth}%
 \MakeFramed {\advance\hsize-\width
   \@totalleftmargin\z@ \linewidth\hsize
   \@setminipage}}%
 {\par\unskip\endMakeFramed%
 \at@end@of@kframe}
\makeatother

\definecolor{shadecolor}{rgb}{.97, .97, .97}
\definecolor{messagecolor}{rgb}{0, 0, 0}
\definecolor{warningcolor}{rgb}{1, 0, 1}
\definecolor{errorcolor}{rgb}{1, 0, 0}
\newenvironment{knitrout}{}{} 

\usepackage{alltt}

\usepackage{verbatim} 
\usepackage{amsthm}
\usepackage{amsbsy}
\usepackage{amssymb}
\usepackage{amsfonts}
\usepackage{graphics}
\usepackage{bm}
\usepackage{mathtools}
\usepackage{bbding}
\usepackage{booktabs}
\usepackage[boxed]{algorithm2e}
\usepackage{hyperref}
\usepackage{algorithmicx}

\usepackage{subfig}

\usepackage{lscape} 
\usepackage{longtable}

\renewcommand{\tablename}{Tab.}

\ifCLASSINFOpdf
\else
\fi

\DeclareMathOperator*{\var}{var}
\DeclareMathOperator*{\tr}{tr}
\DeclareMathOperator*{\cov}{cov}

\DeclareMathOperator*{\argmin}{argmin}

\DeclareMathOperator*{\length}{length}

\DeclareSymbolFont{lettersA}{U}{txmia}{m}{it}
\DeclareMathSymbol{\real}{\mathord}{lettersA}{"92}
\DeclareMathSymbol{\field}{\mathord}{lettersA}{"83}

\IfFileExists{upquote.sty}{\usepackage{upquote}}{}
\begin{document}

\title{A Computationally Efficient Framework for Automatic Inertial Sensor Calibration}

\author{James~Balamuta,~St\'ephane~Guerrier,~Roberto~Molinari,~Wenchao~Yang\thanks{J. Balamuta, S. Guerrier, and W. Yang are with the Department of Statistics, University of Illinois at Urbana-Champaign, 725 S. Wright St., Champaign IL, 61820, USA e-mail: balamut2@illinois.edu, stephane@illinois.edu}
\thanks{R. Molinari is with the Research Center for Statistics, University of Geneva, 1205 Geneva, Switzerland, e-mail: roberto.molinari@unige.ch}
}

\markboth{}%
{Balamuta \MakeLowercase{\textit{et al.}}: A Computationally Efficient Framework for Automatic Inertial Sensor Calibration}

\maketitle

\begin{abstract}
The calibration of (low-cost) inertial sensors has become increasingly important over the past years since their use has grown exponentially in many applications going from unmanned aerial vehicle navigation to 3D-animation. However, this calibration procedure is often quite problematic since the signals issued from these sensors have a complex spectral structure and the methods available to estimate the parameters of these models are either unstable, computationally intensive and/or statistically inconsistent. This paper presents a new software platform for inertial sensor calibration based on the Generalized Method of Wavelet Moments which provides a computationally efficient, flexible, user-friendly and statistically sound tool to estimate and select from a wide range of complex models. In addition, all this is possible also in a robust framework allowing to perform sensor calibration when the data is affected by outliers. The software is developed within the open-source statistical software R and is based on C++ language allowing it to achieve high computational performance.

\end{abstract}


\IEEEpeerreviewmaketitle

\section{Introduction}
An inertial measurement unit (IMU) is a device whose composition is typically a triad of accelerometers and gyroscopes which provide a measurement of the specific force of acceleration as well as the rate of angular movement. With the advances in microtechnology, the premise of a low-cost miniaturized IMU is possible through the construction of a Micro-Electro-Mechanical System (MEMS) device which is a compact and light version of an IMU \cite{el2007promise}. As a result of low cost, light weight and lower energy consumption, the number of applications privy to inertial sensors have increased considerably, leading to a large amount of attention being afforded to device performance by researchers. These developments have thus enabled inertial sensors to be embedded in countless consumer-facing goods that range from smart phones to exercise equipment as autonomous automobiles, while also finding further use in military applications such as unmanned aerial vehicles \cite{woodman2007introduction}.

Despite the considerable amount of applications that MEMS sensors are used for, their small size and low cost compared to the higher grade IMU comes at the price of fairly large and variable measurement errors. As a result, there is a considerable amount of research that solely focuses on modelling and compensating for these errors by attempting to characterize their complex noise structure. To do so, there are different methods which have been regularly used to date, albeit being affected by different drawbacks. The main and most statistically sound of the latter approaches is represented by the Maximum Likelihood (ML) which has been widely studied and used for the task of inertial sensor calibration. In \cite{stebler2011constrained}, the ML was implemented through the use of the Expectation-Maximization (EM) algorithm employed alongside a Kalman filter to attempt the parameter estimation of a State-Space Model (SSM). Although this procedure works well in simple SSM settings, it tends to diverge in only moderately more complex scenarios which are often more realistic for the characterization of inertial sensor error signals. A similar approach was used in \cite{zhao2011stochastic} and \cite{Nikolic2016} to improve the results by implementing, respectively, a nonlinear adaptive Kalman Filter (KF) and a log-sampling scheme without however overcoming the mentioned convergence and numerical instability issues. Moreover, the ML can be time consuming to implement and may require considerable effort to estimate a new model (see \cite{stebler2011constrained}). To mainly tackle the latter obstacle, another existing method is the ``graphical'' Allan Variance Linear Regression (AVLR) which makes use of the slope in the linear regions of a log-log plot of the Allan variance (AV) to estimate the parameters of the underlying processes. For this method, \cite{woodman2007introduction} and \cite{elsheimy08} provide a thorough overview of the approach describing how to distinguish the different underlying processes via graphical representations of the AV. Nevertheless, although still being widely used, \cite{2015GuerrierAVLR} formally proved that this method is inefficient and inconsistent in the vast majority of cases. Considering this, a different use of the AV was presented in \cite{vaccaro2012statistical} where the parameters of a process composed by a white noise and a random walk were simultaneously estimated in a method of moments fashion. 

Given the limitations of existing techniques, a new estimation method was proposed by \cite{guerrier2013wavelet} called the Generalized Method of Wavelet Moments (GMWM). Based on the idea of Generalized Method of Moments (GMM) estimators, the GMWM makes use of the relation between the Wavelet Variance (WV) and the parameters of a process, estimating the latter by minimizing the distance between the empirical and model-based WV (see Section \ref{GMWM}). It must be noted that this method could also be based on the AV since the Haar WV is simply twice the AV. Having said this, in a way this estimator can be viewed as a generalized version of \cite{vaccaro2012statistical} with additional benefits. Among these benefits we can highlight its good statistical properties, its ease of implementation and its computational efficiency. Moreover, the nature of this estimator has allowed to deliver additional statistical tools which are of particular interest for the purpose of inertial sensor calibration. For instance, the development of a model selection criterion to effectively rank models while taking into account their complexity. In addition, there exists a robust estimation version of the GMWM that can be employed when dealing with data that suffers from some form of contamination (e.g. outliers or influential points). Additional research is being developed to extend this estimation framework to multivariate and non-stationary (dynamic) settings which are also of interest for sensor calibration. 

Considering these results and properties, this paper introduces the software platform which allows practitioners and researchers to have wide access to all the tools linked to the GMWM, going from the calculation of the WV to the estimation of parameters and selection of simple and complex error models also in the presence of outliers. In particular, this paper gives details on how these approaches have been implemented through specific algorithms in order for them to be fully tailored to the features which characterize IMU measurement error signals. The platform in which these procedures are delivered can be found in the newly developed ``\texttt{gmwm}'' package within the R open source statistical software and makes use of the C++ language to ensure higher computational efficiency (see Appendix \ref{app:benchmark} for some details). As described in Section \ref{Rpackage}, developing an R package not only allows anyone to access the methodologies and functions of the GMWM but also provides a platform which is efficient and can easily be brought up to date with the most recent developments.

The paper is organized as follows. In Section \ref{GMWM} we give a brief overview of the GMWM and its related existing tools while in Section \ref{Rpackage} we introduce the software platform by presenting its main features in the relative subsections going from model identification to model selection. Finally, Section \ref{upcoming} concludes by giving an overview of the upcoming developments of the GMWM which will be included in future software updates to deliver additional important tools for inertial sensor calibration.

\section{The Generalized Method of Wavelet Moments}
\label{GMWM}

In this section, we provide a brief description of the GMWM estimator along with an overview of the main tools that have been derived and are implemented within the package (for a more thorough description the methods and their properties refer to \cite{guerrier2013wavelet,wvic2015,guerrier2015robust}). As mentioned in the previous section, the GMWM relies on the WV, which is the variance of the wavelet coefficients $(W_{j,t})$, issued from a wavelet decomposition of a signal (see, for example, \cite{percival2000wavelet} for an overview). In the context of the GMWM, this decomposition is currently based on the Haar wavelet and therefore all the results in this paper rely on this type of filter. Furthermore, the platform can provide unbiaed estimations of the WV based on the Discrete Wavelet Transform (DWT) or on the Maximum Overlap Discrete Wavelet Transform (MODWT), with the latter being generally more efficient and therefore being the default method. This estimator is implemented based on the proposal of \cite{Perc:95} who defines the MODWT WV estimator at scale $j$ ($\hat{\nu}_{j}^2$) as
\begin{equation}
 \hat{\nu}_{j}^2 = \frac{1}{T - L_j + 1} \sum_{t=L_j}^{T} {W}_{j,t}^2 .
 \label{eq:MODWTWaveVar}
\end{equation}
where $T$ represents the length of the signal and $L_j$ is the length of the (Haar) wavelet filter at scale $j$. 

Having presented the WV estimator, we can now briefly describe the WV implied by a stochastic process $F_{\bm{\theta}}$ where $\bm{\theta} \in \bm{\Theta} \subseteq \real^p$ represents the parameter vector which defines the model. Denoting $S_{F_{\bm{\theta}}}(f)$ as the Power Spectral Density (PSD) of the process $(Y_t)$, this parameter vector is mapped to the WV through the following relationship
\begin{equation}
 \nu^2_j(\bm{\theta}) = \int_{-1/2}^{1/2} S_{W_j}(f)df = \int_{-1/2}^{1/2} |{H}_j(f)|^2S_{F_{\bm{\theta}}}(f) df .
\label{eq:octavebandestimate}
\end{equation}
where ${H}_j$ is the transfer function of the wavelet filters at scale $j$ and $|\cdot|$ denotes the modulus operator. We can therefore see that the WV is a function of the model parameters thereby providing a theoretical quantity which can be matched with its estimated counterpart. The GMWM is based exactly on the latter idea by attempting to inverse the relation between the WV and the parameters to obtain $\bm{\theta}(\bm{\hat{\nu}})$, where $\bm{\hat{\nu}} = [\nu_j^2]_{j=1,\hdots,J}$. To do so, the GMWM estimator $\hat{\bm{\theta}}$ is defined as
	\begin{equation}
\hat{\bm{\theta}} = \underset{\bm{\theta} \in \bm{\Theta} }{\argmin} \; 
\left(\hat{\bm{\nu}} - \bm{\nu}(\bm{\theta})\right)^{T} \mathbf{\Omega} \left(\hat{\bm{\nu}} - \bm{\nu}(\bm{\theta})\right),
\label{eq:gmwm}
\end{equation}
which therefore is the result of a generalized least squares minimization where $\mathbf{\Omega}$ represents a positive definite weighting matrix chosen in a suitable manner (see e.g. \cite{guerrier2013wavelet}). Conditioned on the positive definiteness of this matrix and denoting the covariance matrix of $\hat{\bm{\nu}}$ as $\bm{V}$, such that $\bm{\nu}\left({\bm{\theta}}\right) \equiv \var\left({\hat{\bm{\nu}}}\right)$, the main role of $\mathbf{\Omega}$ is to make the GMWM estimator as efficient as possible and it was shown that $\mathbf{\Omega} = \bm{V}^{-1}$ is the matrix which allows the GMWM estimator to achieve maximum asymptotic efficiency.

In \cite{guerrier2013wavelet} the large sample properties of the GMWM were studied showing that it is consistent and normally distributed for a wide range of time series models. A robust version of the GMWM was also given in \cite{fastandrobust} allowing to limit the influence of contamination in the observed signal on the estimation process while preserving suitable asymptotic properties. Using these properties and the GMM form of the GMWM, additional useful results were derived for inference purposes allowing to obtain, for example, confidence intervals for the parameters, goodness-of-fit of the estimated models and selection criteria to determine which models are the ``best''.

While the parameter confidence intervals are simply based on the asymptotic normality of the GMWM, the goodness-of-fit test is based on the test for over-identifying restrictions for GMM-type estimators (also known as J-test). The test statistic is given by
\begin{equation}
    T\left(\hat{\bm{\nu}} - \bm{\nu}(\hat{\bm{\theta}})\right)^{T} \mathbf{\Omega}^{*} \left(\hat{\bm{\nu}} - \bm{\nu}(\hat{\bm{\theta}})\right) \xrightarrow[{{H_0}}]{\mathcal{D}} \chi^2_{J-p}
    \label{gof}
\end{equation}
where the $\mathbf{\Omega}^{*}$ denotes an appropriate weighting matrix whose form can be found in \cite{hansen1982large}. The test statistic, as can be seen, follows a $\chi^2$ distribution under the null hypothesis $H_0$ which generally states that the difference between the empirical WV and the WV implied by the estimated model is zero, meaning that the sample moments appear to match the theoretical moments based on $\hat{\bm{\theta}}$. The asymptotic properties associated with this test indicate that the test will not reject $H_0$ asymptotically if the empirical and theoretical moments align. However, it is known that models are only an approximation of reality and, unless there exists a model that is ``true'', the test will nearly always reject this hypothesis when the sample size is extremely large. This is indeed the case for inertial sensor calibration where the length of the measured error signals are in the order of millions. Given this setting, another useful development of the GMWM is the proposal of a selection criterion in \cite{wvic2015} called the Wavelet Information Criterion (WIC) whose estimator is defined as follows
\begin{equation}
    \begin{aligned}
            \text{WI}C &= \underbrace {\mathbb{E}\left[ {{{\left( {\hat{\bm{\nu}}  - \bm{\nu} \left( {\hat{\bm{\theta}} } \right)} \right)}^T}\mathbf{\Omega} \left( {\hat{\bm{\nu}}  - \bm{\nu} \left( {\hat{\bm{\theta}} } \right)} \right)} \right]}_A\\ 
            &+ \underbrace {2\tr\left[ {\cov {{\left[ {\hat{\bm{\nu}} ,\mathbf{\Omega} \bm{\nu} \left( {\hat{\bm{\theta}} } \right)} \right]}^T}} \right]}_B
    \end{aligned}
    \label{eq:wvic.est}
\end{equation}
where $\tr\left[{\cdot}\right]$ denotes the trace operator. This estimator is an unbiased estimate of the ``prediction'' error made by using the estimated parameters from one signal to predict the WV on another signal from the same data-generating process. Term $A$ is sometimes referred to as ``apparent loss'' and is a measure of how well the model fits the observed signal. The latter typically diminishes as the model complexity increases (e.g. adding more underlying models and parameters to the composite process of interest) while term $B$ does the opposite and therefore grows as the model complexity increases. The latter term is sometimes referred to as ``optimism'' and acts as a complexity-based penalty. There are different manners to compute this term such as parametric bootstrap (see \cite{wvic2015}, \cite{efron2004estimation}) or using its asymptotic approximation given in \cite{zhang2015ModelSelection}. Considering these terms, the goal would be to select the model with the smallest WIC value, meaning that we are selecting the model with the smallest estimated prediction error.

To summarize, the GMWM provides a set of extremely useful tools for the task of modelling the purely stochastic errors of inertial sensors for calibration purposes. Based on these theoretical results, the next section presents the new software platform with which it is possible to easily make use of all the advantages of this new approach to signal characterization and modelling.

\section{Implementation of the GMWM}
\label{Rpackage}

The new software platform is developed and made available within the R statistical software which is an open source implementation made available by the R Foundation \cite{rproj}. In this framework, the new platform is implemented within the R package called ``\texttt{gmwm}'' whose main functions have been implemented based on C++ language to ensure a high computational efficiency. This is possible by making use of Rcpp \cite{eddelbuettel2011rcpp} and RcppArmadillo \cite{rcpparmadillo2014} that provide a seamless means to link R to C++ interfaces and the C++ matrix algebra library known as Armadillo \cite{sanderson2010armadillo}. Therefore, having been streamlined in C++, all methods written within the package are highly efficient and are considerably faster than the few existing counterparts implemented in R (see Appendix \ref{app:benchmark}).

Using this flexible and efficient framework, the \texttt{gmwm} package contains all the features that are available within its predecessor,  \cite{GMWMsoftYannick}, while adding a considerable amount of new features and functions. The main features of the new platform can be summarized as follows:
\begin{enumerate}
    \item \textbf{Signal decomposition} across dyadic scales $\tau_{j} = 2^j$, $j = 1,...,J$:
    \begin{itemize}
        \item Supported \textbf{wavelet filters}: Daubechies (d), Fejer-Korovkin (fk), Battle-Lemarie (bl), Least Asymmetric (la), and Minimum Bandwidth (mb)
        \item Discrete Wavelet Transform (\textbf{DWT});
        \item Maximum Overlap DWT (\textbf{MODWT}).
    \end{itemize}
    \item Computation of \textbf{summary statistics}:
        \begin{itemize}
        \item Allan Variance (\textbf{AV}): computed under the traditional definition or the modified definition as described in \cite{riley2008handbook} (denoted as $\sigma _{\bar y}$);
        \item Hadamard Variance (\textbf{HV}): a modification of the AV as described in \cite{riley2008handbook};
        \item Wavelet Variance (\textbf{WV}): computed based on the different filters and transforms (with an option for it to be computed robustly in case of outliers in the signal).
    \end{itemize}
    \item \textbf{Identifying the Models}:
    \begin{itemize}
        \item \textbf{WV plots}: the log-log WV plots are used in the same way as the AV plots where the slopes of the linear parts and the non-linear parts indicate the presence of certain processes in the model which underlies the observed signal (see, for example, \cite{elsheimy08}).
    \end{itemize}
    \item \textbf{Estimating the Models}:
        \begin{itemize}
        \item \textbf{Parameter estimation}: under a supplied model, identified from the WV plot, the ``\texttt{gmwm()}'' function contains many options to estimate the model parameters;
        \item \textbf{Statistical Inference}: either asymptotic or bootstrapped inference is available to create parameter confidence intervals and model goodness-of-fit tests.
    \end{itemize}
    \item \textbf{Model Selection}:
    \begin{itemize}
        \item \textbf{Graphical WV fit}: Having estimated the model, the WV implied by the latter can be compared to the observed WV to understand how well it describes the error signal;
        \item \textbf{Selection procedure}: under a specific set of models (or a general model) a criterion is computed to obtain the ``best'' models for the signal as described in \cite{guerrier2013Algo}.
    \end{itemize}
\end{enumerate}

The first step to take in order to make use of these features is, of course, to load the calibration measurements onto the platform using either the function \texttt{read.imu()} (specifically tailored to loading certain IMU measurements) or the general R function \texttt{read.table()}. In the context of this paper, in order to better describe these features within the following sections, we will make use of real-life IMU data which is made available within a package called ``\texttt{imudata}''\footnote{``\texttt{imudata}'' R package is available at: \url{https://github.com/smac-group/datarepo} or by using ``\texttt{gmwm::install\_imudata()}'' in R}. Each dataset comes from a different sensor (e.g. MTi-G Micro-Electro-Mechanical System (MEMS), NavChip, etc.) and has varying lengths of measurement.  For the purposes of this paper, we focus on examples using the NavChip and MTi-G datasets that contain respectively around $3,100,000$ and $900,000$ \textit{static} observations with six columns representing the measurements for each accelerometer and gyroscope axis. The MTi-G MEMS data was previously analyzed in \cite{guerrier2013wavelet} (we will refer to this dataset as \texttt{mtig}), while the NavChip data is a recent acquire which was sampled at 250Hz for roughly 3.5 hours (we will refer to this dataset as \texttt{navchip}). 

\subsection{Identifying the Models}
\label{wv_view}

\begin{figure*}
  \centering
  \subfloat[WV by Sensor and Axis]{\includegraphics[width=0.48\textwidth]{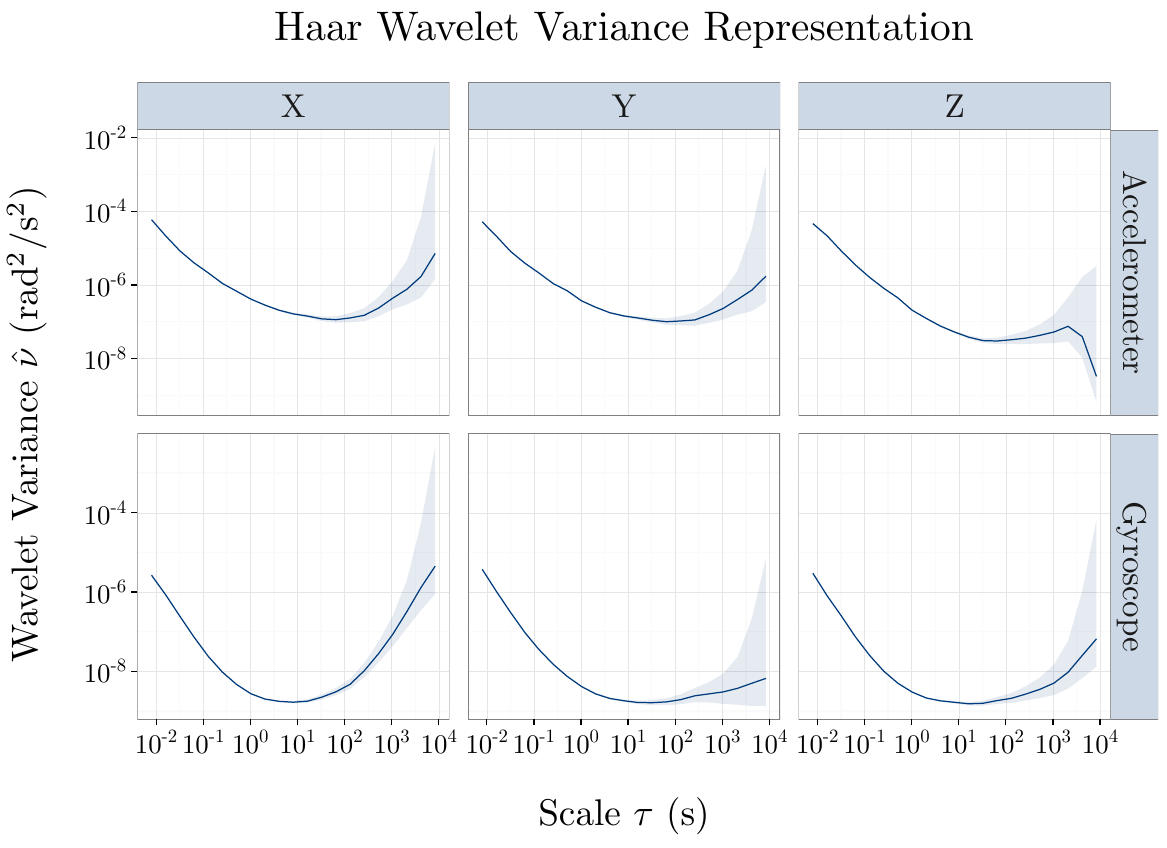}\label{fig:wv_navchip}}
  \hfill
  \subfloat[Overlaid WV by Sensor]{\includegraphics[width=0.48\textwidth]{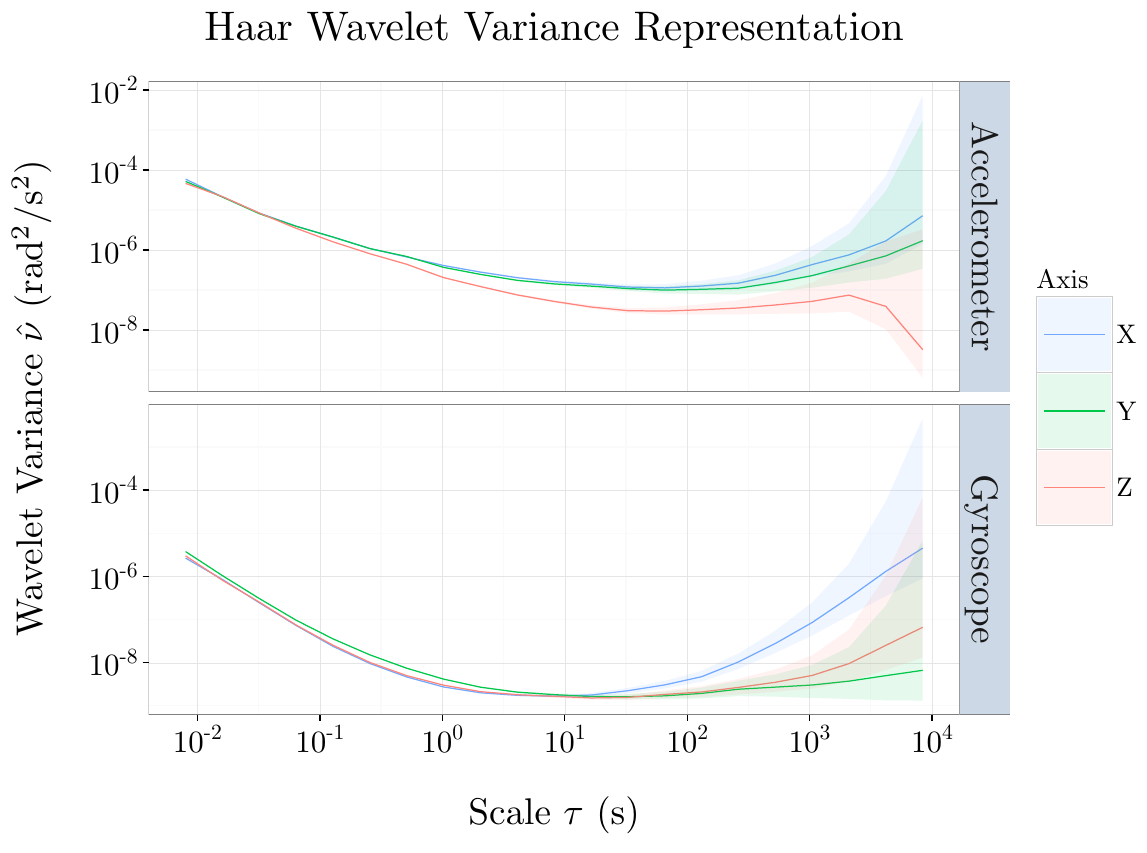}\label{fig:wv_navchip_split}}
  \caption{Plots produced by using the function \texttt{plot()} on the WV issued from the function \texttt{wvar()} on the NavChip IMU. (a) Plot with parameter \texttt{split=TRUE}; (b) Plot with parameter \texttt{split=FALSE}.}
  \label{fig:side-by-side}
\end{figure*}

\begin{figure}
  \centering
  \includegraphics[width=0.48\textwidth]{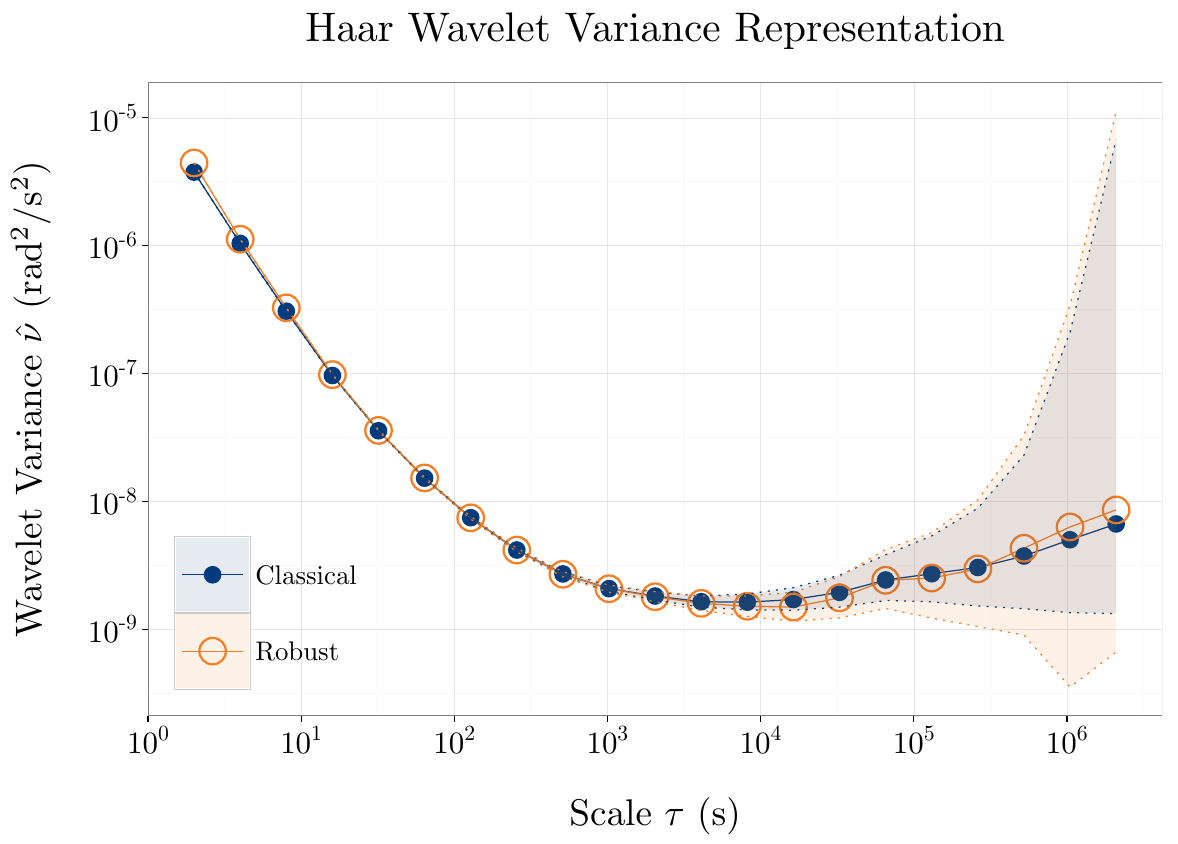}
  \caption{Plot produced by the function \texttt{compare.wvar()} on objects containing the standard and robust WV respectively.}
  \label{fig:comp_wv}
\end{figure}

Once the data is available in the \texttt{R} session (see Appendix \ref{app:imu_load}), we can start the procedure of modelling the stochastic error issued from an inertial sensor which, first of all, consists in understanding what kind of model could best describe the observed signal. This is not necessarily an easy task but a simple and useful tool which is often employed for this purpose is the log-log plot of the AV versus its scales \cite{elsheimy08}. Indeed, according to the form of the AV, a researcher can understand what kind of underlying processes are contributing to the overall error model where a linear or non-linear behaviour in certain regions of the plot can be linked to specific processes. The WV based on the Haar wavelet decomposition is simply a rescaled version of the AV (i.e. $\bm{\sigma}_{\bar y} = 2\bm{\nu}$) and therefore the \texttt{gmwm} package makes use of the same visualization technique (i.e. a log-log plot of the WV versus its corresponding scales). In order to begin the visualization process, one must first obtain the WV of the process(es) by using the ``\texttt{wvar()}'' function. The results of this calculation can then be plotted by using the function ``\texttt{plot()}'' to obtain the plot in Fig. \ref{fig:wv_navchip}, which depicts the WV for each signal separately. By adding the parameter of ``\texttt{split = F}'' to the ``\texttt{plot()}'' function, one is able to have the WV superimposed according to whether the sensor is either a gyroscope or an accelerometer as in Fig. \ref{fig:wv_navchip_split}. 
It is also possible to compute the robust WV (see \cite{guerrier2015robust,mondal2012m}) by adding the ``\texttt{robust = T}'' parameter to the ``\texttt{wvar()}'' function. Within the same function, it is also possible to specify the degree of efficiency of the robust estimator with respect to the standard one where an efficiency close to 0.5 ensures a high level of robustness and viceversa for a value close to 1. Computing the WV from a robust perspective is helpful to check whether there appears to be contamination (e.g. outliers or extreme observations) in the captured signal. Indeed, to understand if modeling within a robust paradigm is required, one should compare the standard WV with the robust version by using the function ``\texttt{compare.wvar()}'' to produce Fig. \ref{fig:comp_wv}.
If the robust WV is visually different from the standard WV, then a robust analysis of the signals is preferable.

\subsection{Estimating the models}
\label{est_model}

The new platform, as seen in the previous section, makes available some flexible plotting tools which are already used (in different forms) to identify the models characterizing inertial sensor stochastic errors. However, the parameters of these models are obviously not known and need to be estimated. As described in Section \ref{GMWM}, the GMWM allows for the estimation of these parameters in an efficient and consistent manner and the function which implements this method is called \texttt{gmwm()}. There are multiple arguments to this function which provide the users with a flexible range of options to tailor the estimation to their needs. In the case of IMUs, there exists a function with preset values ideal to model IMU error signals called \texttt{gmwm.imu()}. Both of these functions rely on users supplying an error model which can be specified using a combination of all or a subset of the following processes:

\begin{itemize}
\item GM($\beta$,$\sigma_{GM}^2$): a Gauss-Markov process;
\item AR1($\phi$,$\sigma^2$): a first-order autoregressive process;
\item WN($\nu^2$): a white noise process;
\item QN($Q^2$): quantization noise process;
\item RW($\gamma^2$): a random walk process;
\item DR($\omega$): a drift process;
\item AR($p$): a $p$-order autoregressive process;
\item MA($q$): a $q$-order moving average process;
\item ARMA($p$, $q$): an autoregressive-moving average process.
\end{itemize}

It must be underlined that a first-order autoregressive process (\texttt{AR1()}) is simply a reparametrization of the Gauss-Markov process (\texttt{GM()}). Denoting the frequency as $f$, the parameterization of the \texttt{AR1()}'s $\phi$ is related to the \texttt{GM()}'s $\beta$ through the following expression
\begin{equation*}
    \phi = \exp{\left({-\beta \Delta t}\right)}
\end{equation*}
where $\Delta t = \frac{1}{f}$. Similarly, the $\sigma^2$ term of the \texttt{AR1()} is related to the \texttt{GM()}'s $\sigma_{GM}^2$ by the expression
\begin{equation*}
    \sigma^2 = \sigma_{GM}^2 \left({1-\exp{\left({-2 \beta \Delta t}\right)}}\right).
\end{equation*}
With this in mind, if the frequency $f$ of the data is equivalent to 1, then the \texttt{GM()} process will be equivalent to the \texttt{AR1()}. This is the default frequency assumed by the software unless otherwise specified during the estimation procedure. 

The latent processes underlying the error signal, whose parameters need to be estimated, can be specified by simply adding the different processes mentioned above via the ``+''  operator. However, there are some limits to how many times a process can be included in a model. In particular, only the \texttt{GM()} or \texttt{AR1()} models can be included more than once (say $k$ times) by specifying, for example, $k$\texttt{*GM()} while the other processes can only be included once within the same model. With these conditions in mind, once the model is specified the software will perform a grid search to obtain appropriate starting values for the optimization procedure in (\ref{eq:gmwm}) (see Appendix \ref{app:init_vals} for details). However, one can also supply starting values for the GMWM optimization by writing the exact values within the bracket (e.g.  ``\texttt{AR1(phi=0.9,sigma2=0.1)+WN(sigma2 = 1)}''). 

To provide an example of model estimation using the above features, let us consider the \texttt{navchip} data and take the second column (Y-axis gyroscope). In order to describe the latter we consider a reasonably complex model made by the sum of three Gauss-Markov processes in addition to a white noise, a quantization noise and random walk process which we estimate as follows:

\begin{verbatim}
    > gyro.y = gmwm.imu(3*GM()+WN()+QN()+RW(), navchip[,2])
\end{verbatim}


As for the \texttt{wvar()} function, also in this case it is possible to opt for a robust model estimation by simply adding the ``\texttt{robust = T}'' parameter. The estimates for both of the models are presented in Tab. \ref{table:gmwm_est} alongside their standard deviation. For the most part, the estimates between the classical and robust GMWM seem to agree since there does not appear to be a significant difference between them. The only exception is represented by the third Gauss-Markov process where $\beta _3$ and $\sigma _{GM,3}^2$ differ between the two estimations. These differences can be also noticed in Fig. \ref{fig:comp_wv} where the standard and robust WV appear to slightly differ (keeping in mind the logarithmic scale). It therefore appears that a robust analysis might be more appropriate.

\begin{table}[!ht] 
\centering
\begingroup\scriptsize
\begin{tabular}{|l|c|c|}
  \cline{2-3}\multicolumn{1}{c}{} &\multicolumn{2}{|c|}{\textbf{Estimates}} \\ 
 \hline
 & Classical & Robust $(\textrm{eff} = 0.6)$ \\ 
  \hline \rule{0pt}{2.6ex}$\beta_{1}$ & $2.50\cdot 10^{-1\phantom{ 1 }}$ ($1.30\cdot 10^{-13}$) & $3.11\cdot 10^{-1\phantom{ 1 }}$ ($1.93\cdot 10^{-13}$) \\ 
   [1.5ex]$\sigma_{GM,1}^2$ & $7.08\cdot 10^{-9\phantom{ 1 }}$ ($5.92\cdot 10^{-13}$) & $6.57\cdot 10^{-9\phantom{ 1 }}$ ($6.58\cdot 10^{-13}$) \\ 
   [1.5ex]$\beta_{2}$ & $6.28\cdot 10^{-3\phantom{ 1 }}$ ($1.83\cdot 10^{-13}$) & $7.26\cdot 10^{-3\phantom{ 1 }}$ ($2.74\cdot 10^{-13}$) \\ 
   [1.5ex]$\sigma_{GM,2}^2$ & $1.28\cdot 10^{-8\phantom{ 1 }}$ ($1.10\cdot 10^{-13}$) & $1.12\cdot 10^{-8\phantom{ 1 }}$ ($1.19\cdot 10^{-13}$) \\ 
   [1.5ex]$\beta_{3}$ & $\bm{8.48\cdot 10^{0\phantom{ -1 }} }$ $\bm{(1.52\cdot 10^{-14})}$ & $\bm{1.25\cdot 10^{1\phantom{ -1 }} }$ $\bm{(3.32\cdot 10^{-14})}$ \\ 
   [1.5ex]$\sigma_{GM,3}^2$ & $6.48\cdot 10^{-9\phantom{ 1 }}$ ($1.44\cdot 10^{-11}$) & $1.33\cdot 10^{-8\phantom{ 1 }}$ ($2.73\cdot 10^{-11}$) \\ 
   [1.5ex]$\sigma^2$ & $6.94\cdot 10^{-7\phantom{ 1 }}$ ($2.01\cdot 10^{-9\phantom{ 1 }}$) & $5.30\cdot 10^{-7\phantom{ 1 }}$ ($2.52\cdot 10^{-9\phantom{ 1 }}$) \\ 
   [1.5ex]$Q^2$ & $2.29\cdot 10^{-6\phantom{ 1 }}$ ($8.79\cdot 10^{-9\phantom{ 1 }}$) & $2.74\cdot 10^{-6\phantom{ 1 }}$ ($1.24\cdot 10^{-8\phantom{ 1 }}$) \\ 
   [1.5ex]$\gamma^2$ & $3.90\cdot 10^{-14}$ ($4.50\cdot 10^{-14}$) & $5.44\cdot 10^{-14}$ ($5.10\cdot 10^{-14}$) \\ 
   \hline
\end{tabular}
\endgroup
\caption{Standard vs robust parameter estimates for the NavChip Y-axis gyroscope. The robust estimator is based on an efficiency of 0.6 compared to the standard estimator.} 
\label{table:gmwm_est}
\end{table}

\begin{figure*}
  \centering
  \subfloat[Comparison of observed WV (blue line) and model-implied WV (orange line)]{\includegraphics[width=0.48\textwidth]{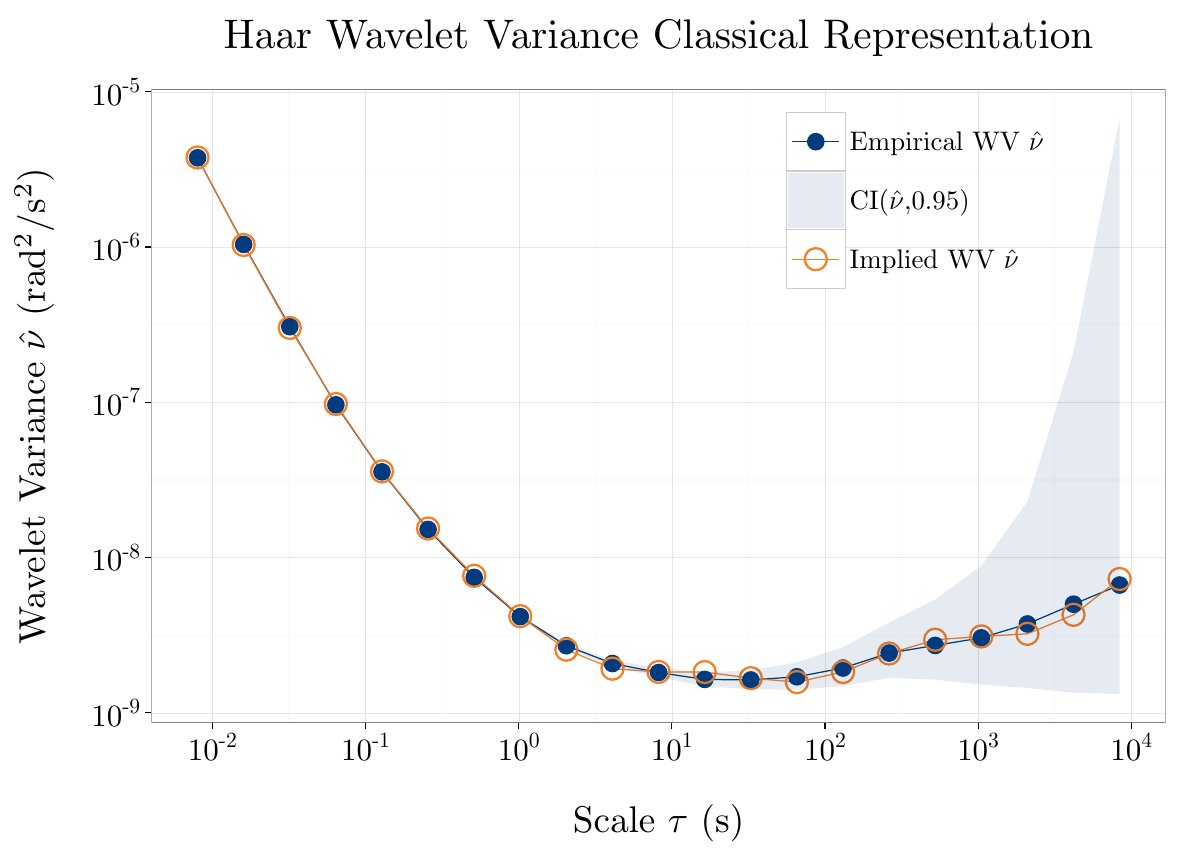}\label{fig:gmwm_nc_gyro_y}}
  \hfill
  \subfloat[WV comparison with underlying process breakdown ]{\includegraphics[width=0.48\textwidth]{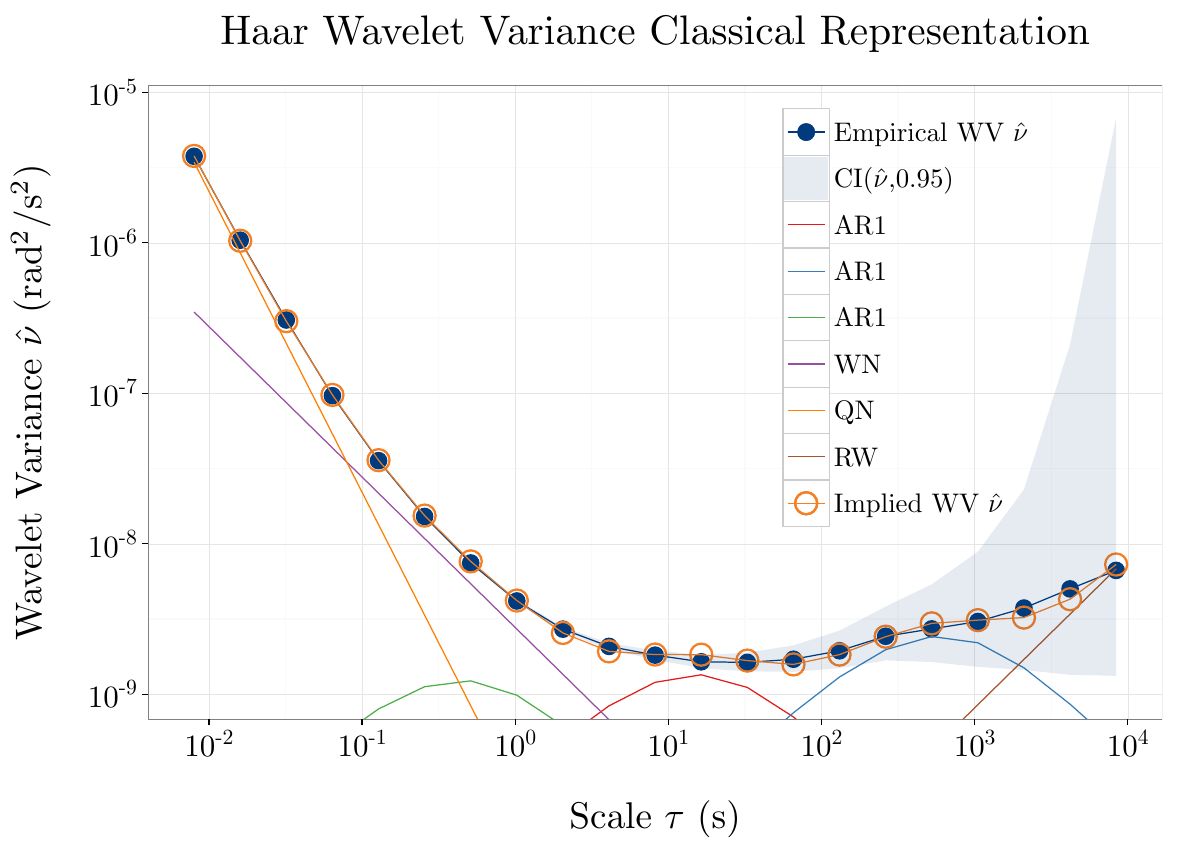}\label{fig:gmwm_nc_gyro_y_decomp}}
  \caption{WV implied by the estimated \texttt{3*AR1()+WN()+QN()+RW()} model compared to the empirical WV of the NavChip Y-axis gyroscope.}
  \label{fig:nc_gyro_y_results}
\end{figure*}

Once the model is estimated, the software provides the tools to graphically validate it. Indeed, a first way to understand if the model fits the observed error signal well is to compare the observed WV and the WV implied by the estimated model as shown in Fig. \ref{fig:gmwm_nc_gyro_y}. This plot can be produced simply by applying the \texttt{plot()} function to the object containing the estimated model. Additionally, one can seek to improve the fit by graphically observing individual processes which contribute to the model by using the parameter ``\texttt{process.decomp = T}'', which gives Fig. \ref{fig:gmwm_nc_gyro_y_decomp}.

Aside from this graphical validation tool, the software provides rigorous statistical measures to assess the significance of the estimated parameters and of the model as a whole. Supposing that the model has been saved in an object called ``\texttt{gyro.y}'', it is possible to obtain parameter confidence intervals (CIs) and a goodness-of-fit test (see equation (\ref{gof})) by using the function ``\texttt{summary(gyro.y, inference = T)}''. The  CIs provide a measure of how accurate the parameters estimations are thereby giving information as to whether or not parameters associated with a specific underlying process are significantly different from zero (e.g. $H_0: \theta = 0$). Thus, this gives a reasonable indication of whether a process should be kept in the model or not. However, with this being said, it is known in the statistics literature that confidence intervals are conditional on the presence of other parameters in the model. As a result, the main conclusion of whether a model is accurate should be obtained from the goodness-of-fit test as it assesses how well the overall model explains the signal.

The tools provided by the new platform and described in this section are extremely important to understand which models can be considered as good candidates to best describe the observed signal. The following step is to select the ``best'' model(s) and the tools provided for this purpose are described in the next section. 

\subsection{Selecting the Model}

\begin{figure}
    \centering
    \includegraphics[width=0.6\textwidth]{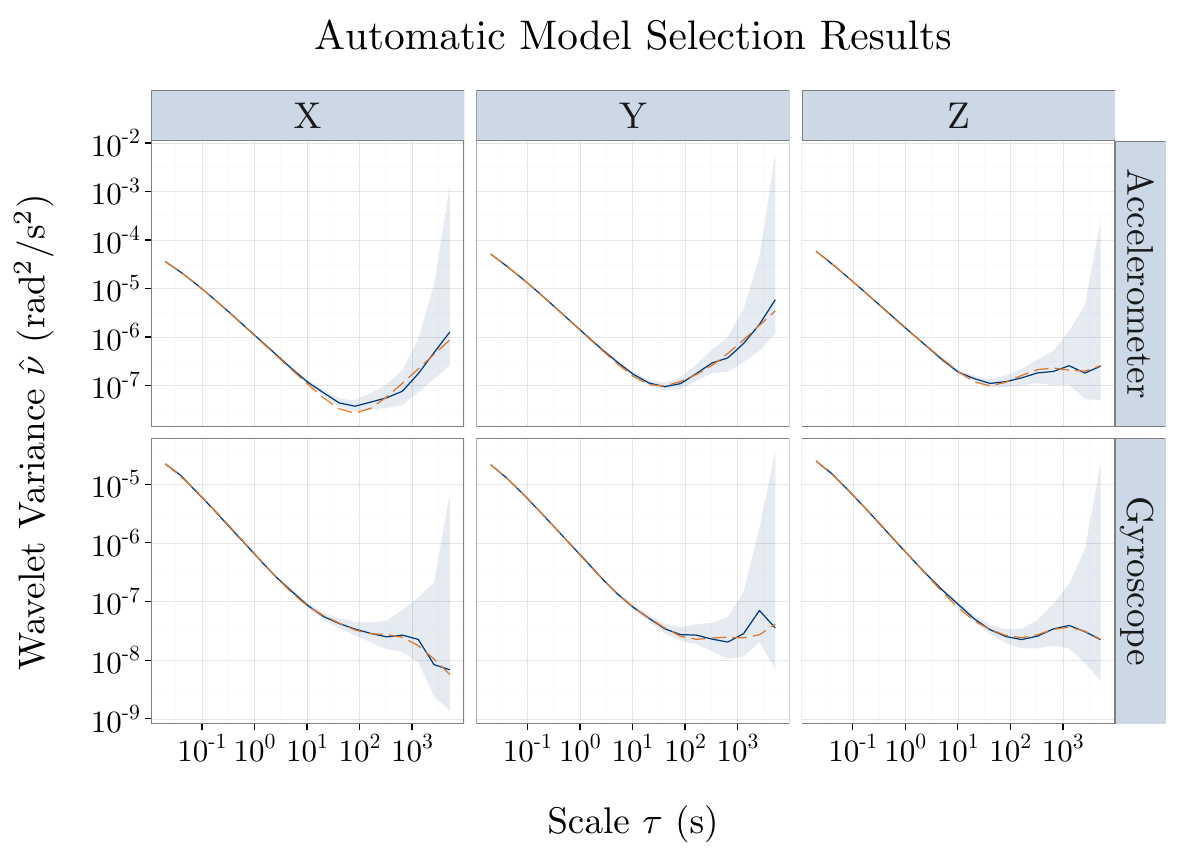}
    \caption{Results of running \texttt{auto.imu()} on the ``\texttt{mtig}'' data. Dashed line is the implied WV while the solid line is the empirical WV.}
    \label{fig:imu6_gmwm}
\end{figure}

The main goal of model selection is to identify the most appropriate model, among those that have been estimated, to describe and predict future stochastic error measurements. Using the WIC, the two main options available to rank the models according to this criterion are the following:
\begin{enumerate}
    \item \textbf{Manual}: the function \texttt{rank.models()} allows to enter a set of candidate models from which the user would like to select the best.
    \item \textbf{Automatic}: the function \texttt{auto.imu()} allows to define an overall general model, say $\mathcal{M}_K$, in which all $K$ candidate models  are nested (e.g. $\mathcal{M}_k \subseteq \mathcal{M}_K$, for $k=1,\hdots,K$). An example of this approach applied to the \texttt{mtig} data is given in Fig. \ref{fig:imu6_gmwm} where the implied WV of the selected models is compared to the observed WV for each accelerometer and gyroscope.
\end{enumerate}

Considering these two options, for both of them the platform also allows to choose the approach to estimate term $B$ in (\ref{eq:wvic.est}) as mentioned in Section \ref{GMWM}. The first option is given by the parametric bootstrap where $H$ simulations are performed on the $k^{th}$ candidate model $F_{\bm{\hat{\theta}}_k}$ to obtain estimates of $\bm{\hat{\nu}}^{(h)}$ and $\bm{\nu}(\bm{\hat{\theta}})_k^{(h)}$ that are then used to estimate the ``optimism'' term. This option has been shown to possess good properties in various simulation studies and can be used by specifying the parameter ``\texttt{bootstrap = T}''. However, the latter is very demanding from a computational standpoint while the asymptotic option, due to its closed analytical form, is less computationally intensive and has some optimal theoretical properties as shown in \cite{zhang2015ModelSelection}. Indeed, in terms of computational time, the asymptotic option is roughly $K$ times faster than the bootstrap option while at the same time being asymptotically loss efficient with a null probability of underfitting.

In conclusion, given the features presented in this section and the previous ones, the challenge of correctly estimating models for inertial sensor stochastic errors can more easily be met through this new platform. The latter not only easily estimates the complex error models which usually characterize IMUs but also delivers additional tools to allow researchers to better understand these error models and to support their decisions for sensor calibration purposes. As a support to the reader and potential user, an example of the code used to generate the results in the previous sections can be found in Appendix \ref{app:code}.

\section{Upcoming features}
\label{upcoming}

The platform responds to the current needs for inertial sensor calibration while adding useful features which were not available before. However, the idea of modelling latent processes which is behind the GMWM allows it to be particularly suitable for extensions to more complicated settings. Different developments are indeed underway, or close to completion, concerning issues which are extremely relevant for inertial sensor calibration. The following sections briefly describe these extensions which will soon be implemented within the new software platform.

\subsection{Multivariate GMWM}
\label{mgmwm}

Currently sensor calibration is carried out separately for each accelerometer and gyroscope that build an IMU. This procedure is valid if one supposes that the error signals of each of these components are independent from each other which may not always be a realistic assumption to make. Using the latent process structure, the idea is to explain the possible dependence of gyroscopes and accelerometers through a common model which is shared among them. To provide an example, we could suppose that the error signals of the triad of gyroscopes $(Y^{(x)}_t,Y^{(y)}_t,Y^{(z)}_t)$
can be expressed as follows:
\begin{equation*}
    \begin{aligned}
        Y^{(x)}_t &= V_t +W^{(x)}_t \\
        Y^{(y)}_t &= V_t +W^{(y)}_t \\
        Y^{(z)}_t &= V_t +W^{(z)}_t
    \end{aligned}
\end{equation*}
where $V_t$ is the process which is shared between the gyroscopes and $W^{(i)}_t$ is the specific process that characterizes the error signal of the i$^{th}$-axis gyroscope. The idea of expressing the model in this manner defines a new method for the modelling of multivariate time series that is inspired by dynamic factor analysis. In this manner, we don't only explain the dependence between the sensor components but we also reduce the number of parameters to estimate and obtain better fits. This feature will be added to the proposed platform once the theoretical properties of the method are confirmed.

\subsection{Calibration under dynamics}
\label{dyngmwm}

The correlation of IMU errors with system dynamics is a topic of great importance that may increase with the spreading use of low-cost inertial sensors. Recent studies confirmed that system dynamics may have an important influence on the characteristics of the error signal issued from inertial sensors (see \cite{stebler2014study}). The possibility of adapting noise models according to the dynamics that the system is subject to should therefore be considered in the navigation filters to enhance the performance of stochastic modeling. The proposed methodology is based on a non-stationary adaptation of the GMWM framework. Let us consider the following example of a non-stationary AR1 process:
\begin{equation*}
    Y_t = \rho_tY_{t-1} + \epsilon_t, \,\,\,\, \epsilon_t \overset{iid}{\sim} (0,\sigma_{\epsilon}^2)
\end{equation*}
where $\rho_t$ is the autoregressive parameter that now depends on the time $t$ through a function $m(\cdot): \mathbb{R} \mapsto \left(-1,1\right)$. Indeed, we define $\rho_t = m(\bm{Z}_{t-1}\beta)$ where $\bm{Z}_{t-1}$ denotes a vector of dynamics and $\beta$ is the parameter that explains the impact of these dynamics on the parameters of the error signal model. This approach can be extended to other parameters of typical error models considered for inertial sensor calibration. The properties of the WV estimator $\bm{\hat{\nu}}$ and the implied WV $\bm{\nu}(\bm{\theta})$ considering such circumstances are being developed and will be later included in the proposed platform.

\subsection{Additional features}
\label{add.feat}

Other features are envisaged for the future updates of the package that are based on recently developed methods and improvements on existing results. For example, it has been shown in \cite{guerrier2013limits} that adding further process moments to the WV vector, such as the the mean or autocovariances, greatly improve the efficiency of the GMWM estimator. This estimator is called the Augmented GMWM (GMWM+) and will be implemented as an additional option in the \texttt{gmwm.imu()} function with the possibility of specifying the requirement on additional moments of the WV vector. Furthermore, a test will be implemented to determine whether two error signals are generated from the same model or not. Finally, additional features are planned such as the possibility to employ other kinds of wavelet filters and use the robust WV as a basis for fault detection and isolation.

\section{Conclusions}
\label{conclusion}

This paper presented the main features and function descriptions of the new software platform for inertial sensor calibration based on the GMWM. This software is contained in an open-source R package called ``\texttt{gmwm}'' which provides all the tools for the modelling of the complex stochastic error signals that characterize IMUs, although this does not exclude the possibility of using it to estimate complex stochastic error models from other engineering applications. As a statistically sound alternative to the AVLR and as a computationally efficient and numerically stable alternative to the ML, the new platform based on C++ language allows to visualize data using the WV as the main summary statistic in order to identify the possible models which can then be quickly and consistently estimated. With graphical tools to assess how well the models fit the observed signals, the package also provides functions to (automatically) select the best model(s) based on their estimated prediction error. These approaches have been implemented in a computationally efficient manner and this paper has also provided details on how these methods have been adapted to the specific and complex characteristics of IMU error signals through tailored algorithms. All this allows to identify the model and estimate its parameters which can then be inserted into a navigation filter (usually an extended Kalman filter) to improve the navigation accuracy of the sensors.

\bibliographystyle{IEEEtran}
\bibliography{main}

\newpage
\appendices

\section{Benchmarks}
\label{app:benchmark}

One of the driving design principles behind the proposed platform is its computational efficiency. To achieve this goal, the computational backend is written completely in C++ using the Armadillo matrix library (see \cite{sanderson2010armadillo}). The implementation of each function is highly efficient due to the manner in which they are implemented. Specifically, multiple functions for the same task were created within C++ and then benchmarked. The benchmarking was conducted using the ``rbenchmark'' package \cite{rbenchmark} with each function being called 100 times. The function that used the least amount of time among the C++ implementations was then included within the package. This bottom-up approach used to create the package not only delivers a very efficient implementation of the GMWM estimator but also of different wavelet-based methods and random process generation. To illustrate, we provide details on the overall computational time across a wide range of sample sizes to estimate a \texttt{3*GM()} model under both the standard and robust settings in two different modes: user-supplied and guessed parameters. The user-supplied parameters were taken to be the exact parameters whereas the guessed parameters relied on the initial starting algorithm described in Appendix \ref{app:init_vals}. The latter algorithm was based on a range of initial guesses going from 100 to 1,000,000. The time reported in Tab. \ref{tab:App1} and \ref{tab:App2} is by the number of \textit{seconds} required to estimate the model.

\begin{table}[!ht]
\centering \footnotesize
\begin{tabular}{rccccc}
  \toprule
Type & 100 & 1,000 & 10,000 & 100,000 & 1,000,000 \\ 
  \midrule
Exact Values & 0.0100 & 0.0188 & 0.0300 & 0.1408 & 1.5473 \\ 
Guessed 1,000 & 0.0342 & 0.0369 & 0.0594 & 0.1714 & 1.5965 \\ 
Guessed 10,000 & 0.0755 & 0.0794 & 0.1250 & 0.2511 & 1.6647 \\ 
Guessed 20,000 & 0.1216 & 0.1386 & 0.2005 & 0.3411 & 1.7917 \\ 
   \bottomrule
\end{tabular}
\caption{Standard estimation of a \texttt{3*GM()} model using ``\texttt{gmwm.imu()}'' in seconds}
\label{tab:App1}
\end{table}

\begin{table}[!ht]
\centering \footnotesize
\begin{tabular}{rcccc}
  \toprule
Type & 1,000 & 10,000 & 100,000 & 1,000,000 \\ 
  \midrule
Exact Values & 0.0327 & 0.0816 & 0.6839 & 11.4097 \\ 
Guessed 1,000 & 0.0546 & 0.1081 & 0.6875 & 11.4575 \\ 
Guessed 10,000 & 0.1135 & 0.1791 & 0.7687 & 11.5385 \\ 
Guessed 20,000 & 0.1666 & 0.2516 & 0.8556 & 11.5957 \\ 
   \bottomrule
\end{tabular}
\caption{Robust estimation of a \texttt{3*GM()} model using ``\texttt{gmwm.imu()}'' in seconds}
\label{tab:App2}
\end{table}

\section{How to Import IMU Data}
\label{app:imu_load}
The first step towards the modelling of the inertial sensor stochastic error is loading the data onto the platform. This is done by: 
\begin{enumerate}
    \item Loading data in \texttt{.txt} or \texttt{.csv} format;
    \item Reading in IMU binary files.
\end{enumerate}
In the first case, the user must also cast the data as an ``\texttt{imu}'' object whereas this is automatically done when loading the data in the second case. Below is an example that displays each of these cases. 

\begin{knitrout}\scriptsize
\definecolor{shadecolor}{rgb}{0.969, 0.969, 0.969}\color{fgcolor}\begin{kframe}
\begin{alltt}
\hlcom{# Case 1a: Reads in data that is separated by tab}
\hlstd{ds} \hlkwb{=} \hlkwd{read.table}\hlstd{(}\hlstr{"path/to/file.txt"}\hlstd{)}

\hlcom{# Case 1b: Reads in data that is separated by comma}
\hlstd{ds} \hlkwb{=} \hlkwd{read.csv}\hlstd{(}\hlstr{"path/to/file.csv"}\hlstd{)}

\hlcom{# Case 2: Cast data loaded into R to IMU object-type}
\hlcom{# for use in the gmwm routines}
\hlstd{sensors} \hlkwb{=} \hlkwd{imu}\hlstd{(ds,} \hlkwc{accelerometer} \hlstd{=} \hlnum{1}\hlopt{:}\hlnum{3}\hlstd{,} \hlkwc{gyroscope} \hlstd{=} \hlnum{1}\hlopt{:}\hlnum{3}\hlstd{,}
    \hlkwc{freq} \hlstd{=} \hlnum{100}\hlstd{)}

\hlcom{# Case 3: Read an imu binary file and cast object as}
\hlcom{# an IMU-type}
\hlstd{sensors} \hlkwb{=} \hlkwd{read.imu}\hlstd{(}\hlstr{"~/path/to/file.imu"}\hlstd{,} \hlstr{"IMU_TYPE"}\hlstd{)}
\end{alltt}
\end{kframe}
\end{knitrout}

The binary file reader currently supports the binary record format used by well known commercial software such as Applanix (PosProc) and Novatel/Waypoint (IExplorer). This format is aptly described as a seven column layout where the first column of the data contains time data, columns 2-4 contain gyroscope values and columns 5-7 contain accelerometer values for axes X, Y, and Z. Upon loading, appropriate scaling factors are applied to the data. As a result, the following types of IMU can be loaded via their binary: IMAR, LN200, LN200IG, IXSEA, NAVCHIP\_INT and NAVCHIP\_FLT. For more details, please see the \texttt{read.imu()} help documentation. 

\section{Optimization Starting Values Algorithm}
\label{app:init_vals}

To start any optimization procedure starting values are needed and this is the case also for the GMWM procedure with the optimization in (\ref{eq:gmwm}). The platform uses an algorithm to find appropriate starting values and this is described in three parts below: a general overview of the algorithm (Algorithm \ref{algo:initial_values}), a heuristic approach to determining the dominating process (Algorithm \ref{algo:dom_values}), and a focus on a specific step based on the required processes to be estimated (Algorithm \ref{algo:process}).

\begin{algorithm}[!htp]
 \SetKwInOut{Input}{Inputs}
  \SetKwInOut{Output}{Output}
  \caption{Starting Values Algorithm}

  \Input{An empty parameter vector ${\bm{\theta}}$ that defines the model, the wavelet variance for the first two scales, and the slope of the data $\left({\frac{{\min \left( x \right) - \max \left( x \right)}}{{\length{(x)}}}}\right)$.}
  \Output{Starting values to estimate the parameter vector $\bm{\theta}$}
 \BlankLine

 \textbf{Step 1}: Compute the total variance $\hat{\sigma}_T^2 = \var{\hat{\bm{\nu}}}$ for all scales obtained with Equation \ref{eq:MODWTWaveVar}.\\
\BlankLine
 \textbf{Step 2}: Determine the process that dominates the initial scales using Algorithm \ref{algo:dom_values}.
 \BlankLine

 \textbf{Step 3a}: Randomly sample $G$ times from the parameter spaces depending on the type of process. At the start of each new round of guessing for all parameters, use the domination result from \textbf{Step 2} to select the starting condition of the ``\texttt{GM}'' / ``\texttt{AR1}'' process:
 
  \uIf{QN dominates}{
    Draw $R \sim U[0,1]$ \\
    \uIf{$R \le 0.75$}{
    Start AR1 or GM on condition 2 in Algorithm \ref{algo:process} 
    }
  }
  \uElseIf{WN dominates}{
    Start AR1 or GM on condition 2 in Algorithm \ref{algo:process}
  }
  
  \BlankLine
 \textbf{Step 3b}: See Algorithm \ref{algo:process} for details on parameter selection for each process.
 \BlankLine

\textbf{Step 4}: Select the starting parameter vector $\hat{\bm{\theta}}^{\left( 0 \right)}$ with the smallest objective function value given by the flattening algorithm described in \cite{yannick2014inf}:

\begin{equation}
    {\hat{\bm{\theta}}^{\left( 0 \right)}} = \mathop {\arg \min }\limits_{\bm{\theta} \in \mathbf{\Theta} } {\left( {1 - \frac{{ \bm{\nu} \left( \bm{\theta} \right)}}{{\hat{\bm{\nu}} }}} \right)^T}\left( {1 - \frac{{\bm{\nu} \left( \bm{\theta}  \right)}}{\hat{\bm{\nu} }}} \right)
\end{equation}

\BlankLine

\label{algo:initial_values}

\end{algorithm}

\begin{algorithm}[!htp]

 \SetKwInOut{Input}{Inputs}
  \SetKwInOut{Output}{Output}
  \caption{Process Domination Algorithm}
    \Input{Process' wavelet variance, and the slope of the data $\left({\frac{{\min \left( x \right) - \max \left( x \right)}}{{\length{(x)}}}}\right)$.}
  \Output{Process that dominates the initial scales.}
 \BlankLine
 
 \textbf{Step 2a}: Compute the slope, $s$, between the first wavelet variance, $\hat{\nu}_1$, and the upper and lower confidence bounds of the second, $a = \left[ {\frac{{\hat \eta {{\hat \nu }_2}}}{{\chi _{\hat \eta }^2\left( {0.975} \right)}},\frac{{\hat \eta {{\hat \nu }_2}}}{{\chi _{\hat \eta }^2\left( {0.025} \right)}}} \right]$, using $ s = \frac{{\log \left( a \right) - \log \left( {{{\hat \nu }_1}} \right)}}{{\log \left( 4 \right)}}$,  where $\hat{\eta} = \max{\left({\left({N - L_j + 1}\right)/2^j,1}\right)}$ 

  \BlankLine
  \textbf{Step 2b}: Determine whether QN, WN, or AR1/GM dominates the initial scales based on a slope heuristic:
  
  \uIf{$\max{(s)} < -.5$}{
    QN Dominates
  }
  \uElseIf{$\min{(s)} > -.5$}{
    AR1/GM Dominates
  }
  \uElse{
    WN Dominates
  }
\BlankLine

\label{algo:dom_values}

\end{algorithm}

\begin{algorithm}[!htp]
 \SetKwInOut{Input}{Inputs}
  \SetKwInOut{Output}{Output}
  \caption{Process-specific Starting Values Algorithm}

  \Input{An empty parameter vector ${\bm{\bm{\theta}}}$ that defines the model, the process' wavelet variance, and the slope of the data.}
  \Output{A parametric model $F_{\bm{\bm{\theta}}}$ }
 \BlankLine

 \textbf{Step 3b}: Select the appropriate realization generator based on process type \\
  \uIf{AR1 or GM}{
      \uIf{AR/GM on condition 1}{
        Draw $U \sim U\left({0,1/3}\right)$ for $\phi = {\frac{{1 - \sqrt {1 - 3U} }}{5}}$  \\
        Draw ${\sigma ^2}\sim U\left( {\frac{{\widehat \sigma _T^2{{\left( {1 - \phi } \right)}^2}}}{2},\widehat \sigma _T^2{{\left( {1 - \phi } \right)}^2}} \right)$
      }
      \uElseIf{AR/GM on condition 2}{
        Draw ${\phi _i}\sim U\left( {\max \left( {0.9,{\phi _{i - 1}}} \right),{\text{0.999995}}} \right)$ \\
        Draw ${\sigma ^2} \sim U\left( {0,\frac{{\hat \sigma _T^2\left( {1 - \phi _i^2} \right)}}{{100}}} \right)$
      }
      \uElse{
        Draw $U \sim U\left({0,1/3}\right)$ for ${\phi _i} = \left( {.999995 - {\phi _{i - 1}}} \right)\left( {\sqrt {1 - 3U} } \right) + {\phi _{i - 1}}$ \\
        Draw ${\sigma ^2}\sim U\left( {\frac{{\widehat \sigma _T^2{{\left( {1 - \phi } \right)}^2}}}{2},\widehat \sigma _T^2{{\left( {1 - \phi } \right)}^2}} \right)$ 
      }
      Increase the condition by 1.
  }
  \uElseIf{DR}{
   Calculate the slope of the data: $ R = \frac{{\max \left( x \right) - \min \left( x \right)}}{{\length\left( x \right)}}$ \\
   Draw $\omega \sim U\left( {\frac{R}{{100}},\frac{R}{2}} \right)$
  }
  \uElseIf{RW}{
    Draw ${\gamma ^2} \sim U\left( {\frac{{\hat \sigma _T^2}}{{{{10}^5} \cdot N}},\frac{{\hat \sigma _T^2}}{N}} \right)$
  }
  \uElseIf{WN}{
    \uIf{WN Dominates}{
    Draw ${\sigma ^2}\sim U\left( {\frac{{\hat \sigma _T^2}}{2},\hat \sigma _T^2} \right)$
    }
    \uElse{
    Draw ${\sigma ^2}\sim U\left( {\frac{{\hat \sigma _T^2}}{{{{10}^5}}},\frac{{\hat \sigma _T^2}}{{10}}} \right)$
    }
  }
  \uElse{
    \uIf{QN Dominates}{
    Draw ${Q^2}\sim U\left( {\frac{{\hat \sigma _T^2}}{8},\frac{{\hat \sigma _T^2}}{3}} \right)$
    }
    \uElse{
    Draw ${Q^2}\sim U\left( {\frac{{\hat \sigma _T^2}}{{2 \cdot {{10}^5}}},\frac{{\hat \sigma _T^2}}{{100}}} \right)$

    }
  }

\label{algo:process}

\end{algorithm}

\newpage

\section{Code Used in Paper}
\label{app:code}

The following code was used to generate some of the results and figures included in the paper and provides insight to the use of the ``\texttt{gmwm}'' platform. Please note that the following code was executed under version 3.0.0 of both the ``\texttt{gmwm}'' and ``\texttt{imudata}'' R package.

\begin{knitrout}\scriptsize
\definecolor{shadecolor}{rgb}{0.969, 0.969, 0.969}\color{fgcolor}\begin{kframe}
\begin{alltt}
\hlcom{# 0. Preparing the workspace}
\hlcom{# Load R Package}
\hlkwd{library}\hlstd{(}\hlstr{"gmwm"}\hlstd{)}

\hlcom{# Load in MTIG data. If the data package is not installed,}
\hlcom{# then install it.}
\hlkwa{if}\hlstd{(}\hlopt{!}\hlkwd{require}\hlstd{(}\hlstr{"imudata"}\hlstd{))\{}
  \hlkwd{install_imudata}\hlstd{()}
  \hlkwd{library}\hlstd{(}\hlstr{"imudata"}\hlstd{)}
\hlstd{\}}

\hlcom{# Read in Binary IMU Data}
\hlstd{navchip} \hlkwb{=} \hlkwd{read.imu}\hlstd{(}\hlstr{"~/IMU_data/navchip_4h.imu"}\hlstd{,}\hlstr{"NAVCHIP_FLT"}\hlstd{)}

\hlcom{# 1.  WAV-AR All Navchip}
\hlcom{# Obtain the wavelet variance of all sensors }
\hlcom{# in imu object.}
\hlstd{wv.navchip} \hlkwb{=} \hlkwd{wvar}\hlstd{(navchip)}

\hlcom{# Plot the Wavelet Variance of each sensor}
\hlkwd{plot}\hlstd{(wv.navchip)}

\hlcom{# 2. Generate models for each sensor}
\hlstd{navchip.gx} \hlkwb{=} \hlkwd{gmwm.imu}\hlstd{(}\hlnum{2}\hlopt{*}\hlkwd{GM}\hlstd{()} \hlopt{+} \hlkwd{WN}\hlstd{()} \hlopt{+} \hlkwd{DR}\hlstd{()} \hlopt{+} \hlkwd{QN}\hlstd{()} \hlopt{+} \hlkwd{RW}\hlstd{(),}
                      \hlstd{navchip[,}\hlnum{1}\hlstd{])}

\hlstd{navchip.gy} \hlkwb{=} \hlkwd{gmwm.imu}\hlstd{(}\hlnum{3}\hlopt{*}\hlkwd{GM}\hlstd{()} \hlopt{+} \hlkwd{WN}\hlstd{()} \hlopt{+} \hlkwd{QN}\hlstd{()} \hlopt{+} \hlkwd{RW}\hlstd{(),}
                      \hlstd{navchip[,}\hlnum{2}\hlstd{])}

\hlstd{navchip.gz} \hlkwb{=} \hlkwd{gmwm.imu}\hlstd{(}\hlnum{2}\hlopt{*}\hlkwd{GM}\hlstd{()} \hlopt{+} \hlkwd{WN}\hlstd{()} \hlopt{+} \hlkwd{DR}\hlstd{()} \hlopt{+} \hlkwd{QN}\hlstd{()} \hlopt{+} \hlkwd{RW}\hlstd{(),}
                      \hlstd{navchip[,}\hlnum{3}\hlstd{])}

\hlstd{navchip.ax} \hlkwb{=} \hlkwd{gmwm.imu}\hlstd{(}\hlnum{4}\hlopt{*}\hlkwd{GM}\hlstd{()} \hlopt{+} \hlkwd{QN}\hlstd{()} \hlopt{+} \hlkwd{RW}\hlstd{(), navchip[,}\hlnum{4}\hlstd{])}

\hlstd{navchip.ay} \hlkwb{=} \hlkwd{gmwm.imu}\hlstd{(}\hlnum{4}\hlopt{*}\hlkwd{GM}\hlstd{()} \hlopt{+} \hlkwd{QN}\hlstd{()} \hlopt{+} \hlkwd{RW}\hlstd{(), navchip[,}\hlnum{5}\hlstd{])}

\hlstd{navchip.az} \hlkwb{=} \hlkwd{gmwm.imu}\hlstd{(}\hlnum{3}\hlopt{*}\hlkwd{GM}\hlstd{()} \hlopt{+} \hlkwd{QN}\hlstd{(),  navchip[,}\hlnum{6}\hlstd{])}

\hlcom{# Robust model generation}
\hlstd{navchip.gy.r} \hlkwb{=} \hlkwd{gmwm.imu}\hlstd{(}\hlnum{3}\hlopt{*}\hlkwd{GM}\hlstd{()} \hlopt{+} \hlkwd{WN}\hlstd{()} \hlopt{+} \hlkwd{QN}\hlstd{()} \hlopt{+} \hlkwd{RW}\hlstd{(),}
                        \hlstd{navchip[,}\hlnum{2}\hlstd{],} \hlkwc{robust} \hlstd{= T)}

\hlcom{# Summary statistics}
\hlstd{sm.navchip.gy} \hlkwb{=} \hlkwd{summary}\hlstd{(navchip.gy,} \hlkwc{inference} \hlstd{= T)}
\hlstd{sm.navchip.gy.r} \hlkwb{=} \hlkwd{summary}\hlstd{(navchip.gy.r,} \hlkwc{inference} \hlstd{= T)}

\hlcom{# 3. Compare Classical with Robust WV}
\hlkwd{compare.wvar}\hlstd{(navchip.gy, navchip.gy.r)}

\hlcom{# 4. Load and cast data for MTiG}
\hlkwd{data}\hlstd{(}\hlstr{"imu6"}\hlstd{)}

\hlcom{# Cast as an IMU}
\hlstd{sensors} \hlkwb{=} \hlkwd{imu}\hlstd{(imu6,} \hlkwc{gyros} \hlstd{=} \hlnum{1}\hlopt{:}\hlnum{3}\hlstd{,} \hlkwc{accels} \hlstd{=} \hlnum{4}\hlopt{:}\hlnum{6}\hlstd{,} \hlkwc{freq} \hlstd{=} \hlnum{100}\hlstd{)}

\hlcom{# 5. Obtain WV and plot WV}
\hlstd{wv.sensors} \hlkwb{=} \hlkwd{wvar}\hlstd{(sensors)}

\hlkwd{plot}\hlstd{(wv.sensors)}

\hlcom{# 6. Automatic Model Selection with Overall Model Guess}
\hlstd{models.sensors} \hlkwb{=} \hlkwd{auto.imu}\hlstd{(sensors,}
                          \hlkwc{model} \hlstd{=} \hlnum{4} \hlopt{*} \hlkwd{AR1}\hlstd{()} \hlopt{+} \hlkwd{WN}\hlstd{()} \hlopt{+} \hlkwd{RW}\hlstd{())}

\hlcom{# 7. Visually see the best models}
\hlkwd{plot}\hlstd{(models.sensors)}
\end{alltt}
\end{kframe}
\end{knitrout}

\end{document}